\documentclass[aps,twocolumn,showpacs,floatfix]{revtex4}

\usepackage{amsfonts}
\usepackage{amsmath}
\usepackage{amssymb}
\usepackage{graphicx}
\usepackage{color}
\usepackage{verbatim}
\usepackage{lipsum}
\usepackage{dsfont}

\usepackage[T1]{fontenc}

\usepackage{hyperref}
\usepackage{makecell}

\begin{document}

\title{Cross-relaxation studies with optically detected magnetic resonances in nitrogen-vacancy centers in diamond in an external magnetic field}
\author{Reinis Lazda$^1$}
\email{reinis.lazda@lu.lv}
\author{Laima Busaite$^1$}
\email{laima.busaite@lu.lv}
\author{Andris Berzins$^1$}
\author{Janis Smits$^1$}
\author{Marcis Auzinsh$^1$}
\author{Dmitry Budker$^{2,3,4}$}
\author{Ruvin Ferber$^1$}
\author{Florian Gahbauer$^1$}

\affiliation{$^1$Laser Centre, University of Latvia}
\affiliation{$^2$Helmholtz-Institut, GSI Helmholtzzentrum f{\"u}r Schwerionenforschung, 55128 Mainz, Germany}
\affiliation{$^3$Johannes Gutenberg-Universit{\"a}t Mainz, 55128 Mainz, Germany}
\affiliation{$^4$Department of Physics, University of California at Berkeley, USA}


\begin{abstract}

In this paper cross-relaxation between nitrogen-vacancy (NV) centers and substitutional nitrogen in a diamond crystal was studied. It was demonstrated that optically detected magnetic resonance signals (ODMR) can be used to measure these signals successfully. The ODMR were detected at axial magnetic field values around 51.2~mT in a diamond sample with a relatively high (200~ppm) nitrogen concentration.  We observed transitions that involve magnetic sublevels that are split by the hyperfine interaction. Microwaves in the frequency ranges from 1.3 GHz to 1.6 GHz ($m_S=0\longrightarrow m_S=-1$ NV transitions) and from 4.1 to 4.6 GHz ($m_S=0\longrightarrow m_S=+1$ NV transitions) were used.

To understand the cross-relaxation process in more detail and, as a result, reproduce measured signals more accurately, a model was developed that describes the microwave-initiated transitions between hyperfine levels of the NV center that are undergoing anti-crossing and are strongly mixed in the applied magnetic field. Additionally, we simulated the extent to which the microwave radiation used to induce ODMR in the NV center also induced transitions in the substitutional nitrogen via cross-relaxation.

The improved understanding of the NV processes in the presence of a magnetic field will be useful for designing NV-diamond-based devices for a wide range of applications from implementation of q-bits to hyperpolarization of large molecules to various quantum technological applications such as field sensors.

\end{abstract}

\maketitle

\section{Introduction}

Negatively charged nitrogen-vacancy (NV) color centers in diamond crystals currently are considered as excellent candidates for a very broad range of applications. Ranging from q-bits implemented on NV centers~\cite{Popkin1,Ladd} and quantum memory elements~\cite{Heshami1} for quantum computers, to probes for different physical fields like magnetic field~\cite{Zheng1,Taylor}, electric field~\cite{Dolde}, and temperature~\cite{Clevenson1,Acosta}. NV centers could serve as tools for achieving hyperpolarization of a variety of molecules as well~\cite{Plenio1,Pines1,Broadway1}. NV centers in diamond can also be used as probes for electron spin and nuclear spin resonance measurements of spins and nanoscale magnetic structures attached to the surface of a diamond crystal~\cite{Mamin1, Kolkowitz1, Zhao1, Bud1, Bud2, Wood1, Wang1,Staudacher1}. The recent progress in the studies of NV centers in diamond suggests that very soon these studies will move from the stage of quantum physics and quantum information science to a stage where they will be implemented in quantum technologies~\cite{Aharonovich1}.



Inside the diamond crystals exist a wide range of paramagnetic defects. Among these defects are substitutional nitrogen (P1 center), which occurs when one of the $^{12}\text{C}$ atoms is replaced by a nitrogen atom $^{14}\text{N}$, or point-like defects, which can occur, for example, when one of the carbon $^{12}\text{C}$ isotopes is replaced by a carbon $^{13}\text{C}$. NV centers can be used as efficient sensors that are already inside the diamond crystal to study these defects~\cite{Wyk1,Ajoy1,Shenderova1}, which are interesting not only by themselves but for the  role they play in the hyperpolarization effects of the nuclear spin of nitrogen. Initial hyperpolarization of the nitrogen nucleus through optically pumped NV electrons have a broad range of applications~\cite{Schwartz}, but can be influenced by other point-like defects in the diamond crystal lattice~\cite{Ajoy2}.


In the study reported in this paper we examine cross-relaxation processes -- energy exchange between hyperpolarized NV centers and substitutional nitrogen (P1 centers) in a diamond lattice when the energy splitting of the magnetic sublevels in the NV center matches the magnetic sublevel splitting in the P1 center at a particular value of the applied external magnetic field $\bf B$. The substitutional nitrogen has an electron spin $S = 1/2$. In our study we use the negatively charged NV centers that have an extra electron combined with one electron from the vacancy, which results in an $S=1$ system called NV$^-$, denoted NV in this work. NV$^0$ centers with an unpaired electron exist in diamonds as well, but will not be considered in this study.

Changes caused by cross-relaxation in optically polarized NV centers can be measured by means of optically detected magnetic resonances (ODMR)~\cite{Negyedi1}. The properties of these magnetic resonances that make them especially suited for cross-relaxation studies include spin-polarization-dependent fluorescence intensity, the coupling of the electron spin to the magnetic environment, and a spin polarization lifetime that is long compared to other similar systems, even at room temperature~\cite{Doherty1}.

Magnetic dipole transitions, caused by an applied microwave (MW) field, between spin magnetic sublevels in NV centers can be used to monitor the electron spin polarization of the NV centers and to study processes such as cross-relaxation that influence the electron spin polarization.

The advantage of using a continuous-wave (CW) ODMR experiment in comparison to a pulsed experiment (for example,~\cite{Bud2}) is that there is no need for synchronised pulse setups. Also one can use lower MW power, and MW homogeneity does not play such an important role in CW ODMR experiments as in the pulsed field experiments.



\section{Cross-relaxation process}


Because NV centers can be optically initialized and read out, they can be used to study other paramagnetic defects in the local environment, as well as interactions between defects. Here, we investigate the dynamics of cross-relaxation between NV and P1 centers.


There exist two families of P1 centers in the diamond crystal depending on whether the symmetry axis of the unpaired electron's orbital is parallel to the NV center's axis ("on-axis"), which is also the direction of the applied magnetic field, or along one of the other three possible axes in the diamond crystal that are not aligned along the external magnetic field ("off-axis")~\cite{Wood1}. It is also important to remember that each of the P1 centers switches between all four symmetry axes on a time scale of a few milliseconds~\cite{Hanson1,xiao_proposal_2016}. This time scale is much shorter than our measurement time, and as a result we see signals corresponding to all possible orientations simultaneously for every P1 center. This hints us that on-axis cross-relaxation peaks could be roughly three times weaker than the off-axis cross-relaxation peaks.




\begin{figure}[h!]
  \begin{center}
    \includegraphics[width=0.45\textwidth]{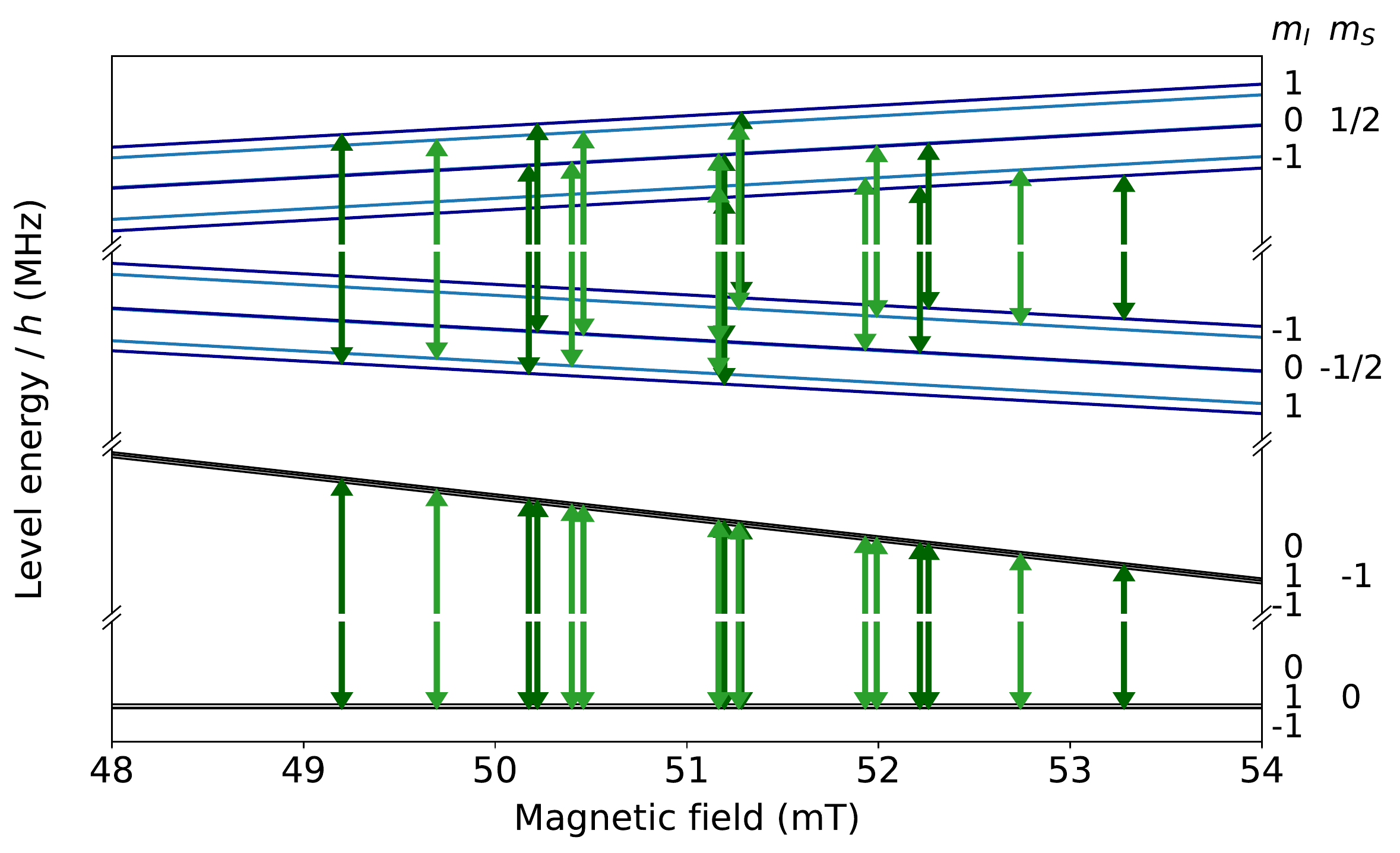}
  \end{center}
  \caption{Level diagram for NV (black) and P1 (blue) centers with hyperfine interaction in the vicinity of $51.2~\mathrm{mT}$. The arrows show the energy-matching transitions between the NV center's $\vert m_S=0,m_I\rangle$ and $\vert m_S=-1,m_I\rangle$ magnetic sublevels, and the P1 center's $\vert m_S=-\frac{1}{2},m_I\rangle$ and $\vert m_S=\frac{1}{2},m_I\rangle$ magnetic sublevels. Each of the arrows are split into three by the NV center's hyperfine structure, but they are omitted in the figure for better readability.}
  \label{energy_levels}
\end{figure}

As discussed already in 1959.~\cite{Bloembergen1}, the angular momentum of a spin system is not generally conserved in cross-relaxation processes. Similar to the Barnett and Einstein-de Haas effects, the balance of the angular momentum is transferred to the rigid lattice. 
Cross-relaxation has been experimentally observed in a variety of spin systems~\cite{Wang1, Holliday1, Oort1, Hiromitsu1, Afach1}. There are no strong selection rules for electron spin $S$ projection changes $\Delta m_S$ or nuclear spin $I$ projection changes $\Delta m_I$ in cross-relaxation processes. The only requirement is that there is an energy matching between the energy-level splitting in the two sub-ensembles~\cite{Wang1}, in our case the NV centers and the P1 centers.


In this situation the magnetic dipole--dipole interaction~\cite{Wood1} couples the two transitions (one of the NV center's states with one of the P1 center's states) and leads to an efficient energy exchange between these two systems (cross-relaxation process). This energy exchange leads to an increase in the effective relaxation rate of the NV center~\cite{Wood1,Bud2}. One can expect that, as a result, the contrast of the ODMR signals will be decreased, and the width of these resonances will be increased. Both parameters can be experimentally measured and these are the effects that we are investigating in this paper.

Besides the electron spin $S = 1/2$ due to the nitrogen valence electron, the P1 center has a nuclear spin $I = 1$ due to the nitrogen nucleus. However, these spins are decoupled by the internal interactions in the diamond crystal and by the applied external magnetic field $\bf B$. Likewise, the NV center, in addition to the electron spin $S = 1$, has a nuclear spin $I = 1$ due to the nitrogen atom of the NV center. These spins are decoupled for the same reasons as in the P1 center. At magnetic field values around $51.2$~mT the energy difference between the decoupled hyperfine-structure magnetic sublevels for the P1 center $\vert m_S=+1/2, m_I\rangle$ and $\vert m_S=-1/2, m_I\rangle$ coincides with the energy difference between the decoupled hyperfine magnetic sublevels for the NV center $\vert m_S=-1, m_I\rangle$ and $\vert m_S=0, m_I\rangle$, and cross-relaxation can occur. As the nuclear gyromagnetic ratio in nitrogen $\gamma_{^{14}\text{N}}$ is much smaller than the electron gyromagnetic ratio $\gamma_{e}$, let us neglect the nuclear spin for the moment.

At zero magnetic field the spin projection states for the P1 center electron spin states $\vert m_S= +1/2 \rangle$ and $\vert m_S= -1/2 \rangle$ have equal energy, but the spin states for the NV center $\vert m_S=0\rangle$ and $\vert m_S=\pm 1\rangle$ are split by the zero-field splitting, which is approximately equal to $2.87$~GHz \cite{Doherty1}, see FIG. \ref{levels}. Since we neglect nuclear spin, both centers are pure electron spin states and have the same magnetic momentum, approximately equal to two Bohr magnetons $\gamma_{e} \approx 2\mu_B$. With these parameters the energy splittings of the electron spin projection states described above have equal values around the magnetic field value of 51.2~mT, see FIG. \ref{energy_levels}.

\section{ODMR method}

\begin{figure}[h!]
  \begin{center}
    \includegraphics[width=0.4\textwidth]{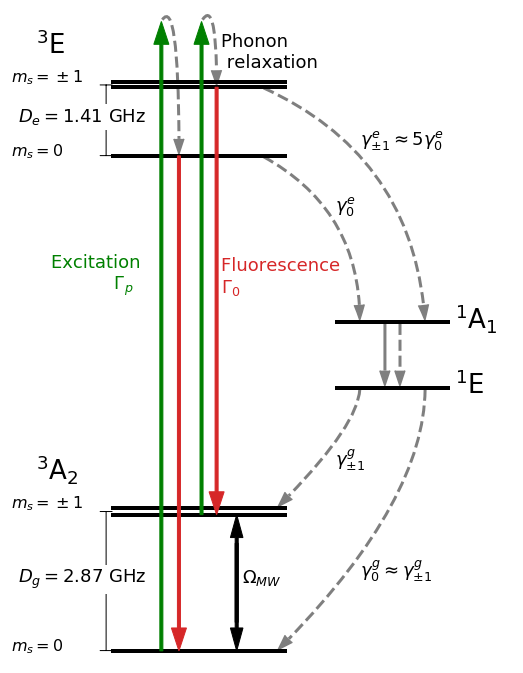}
  \end{center}
  \caption{Level scheme of an NV center in diamond, $m_S$ is the electron spin-projection quantum number, $D_g$ and $D_e$ are the ground-state and excited-state zero-magnetic-field splittings, $\Omega_{MW}$ is the MW Rabi frequency, $\gamma_0^g$ and $\gamma_{\pm 1}^g$ are the relaxation rates from the singlet state $^1$E to the triplet ground state $^3$A$_2$, $\gamma_0^e$ and $\gamma_{\pm 1}^e$ are the relaxation rates from the triplet excited state $^3$E to the singlet state $^1$A$_1$ \cite{Auzinsh1}.
  }
  \label{levels}
\end{figure}

The ODMR method can be described using the NV center energy-level scheme shown in FIG.~\ref{levels}.
Green laser light was used to excite optically the NV centers. In the model we assume that the excitation transitions are spin-projection conserving, $\Delta m_S=0$, irrespective of the laser light polarization. We also assumed that the absorption rates for different laser polarizations and different transitions are equal for all three spin-projection states, $\Gamma_p$ in FIG.~\ref{levels}).
After light absorption and rapid relaxation in the phonon band, the excited-state magnetic sublevels either can decay back to the ground state with equal rates $\Gamma_0$ and radiate light with a wavelength in the red part of the electromagnetic radiation spectrum or undergo non-radiative transitions to the singlet level $^{1}A_1$ (see FIG.~\ref{levels}). These non-radiative transitions occur roughly five times more rapidly from the excited state electron spin sublevels $m_S = \pm 1$ than from $m_S = 0$. After that the singlet-singlet transition ${^{1}A}_1 \longrightarrow$ $^{1}E$ takes place with almost all the energy being transferred in a non-radiative way with a small portion being IR radiation. And finally, non-radiative transitions from $^{1}E$ to the triplet ground state occur with roughly equal transition probabilities to all three electron spin projection components of the $^3A_2$ level. It can be understood that the differences between the non-radiative transition rates in the excited triplet state $^3E$ lead to the situation that after several excitation-relaxation cycles the population of the NV centers in the triplet ground state will be transferred to the magnetic sublevel $m_S = 0$, and the electron spin angular momentum will be strongly polarized~\cite{Doherty1}.

This electron-spin polarization in the ground state of the NV centers leads to an increase in the red fluorescence intensity. If in addition to the laser radiation, MW radiation is applied in resonance to a certain transition from the $m_S = 0$ spin polarized state to one of the $m_S = \pm 1$ states, the population of NV centers in the $m_S=0$ state is reduced, and hence, the fluorescence intensity diminishes. This effect is the essence of the ODMR method.

By analyzing the dependence of the experimentally measured ODMR signal contrast and width on the external magnetic field we studied NV and P1 center cross-relaxation at magnetic field values around 51.2~mT.

\section{Hyperfine splitting, hyperfine level mixing and nuclear spin polarization}
\label{signals}

Measurements and signal analysis are complicated by the fact that each of the P1 and NV spin magnetic sublevels in diamond are split into three components by the hyperfine interaction of electron spin with the nuclear spin ($I=1$) of nitrogen $^{14}{\rm N}$. This hyperfine interaction is substantially smaller in the NV center than in the P1 center, see Eq.~\eqref{eq1}. As a result, in our experiment with a diamond sample of 200~ppm $^{14}{\rm N}$ concentration in cross-relaxation transitions involving the individual hyperfine components of the P1 centers the signals can be resolved, but individual hyperfine components of NV centers cannot be resolved, see FIG.~\ref{energy_levels} as well as FIG.~\ref{minus1} and FIG.~\ref{plus1} in section~\ref{label1}.




For the analysis of the obtained results we look at two Hamiltonians. The first one describes the NV center in the ground state and includes the zero-field splitting, the hyperfine interaction in the NV center, and the interaction of the NV center with an external magnetic field. We find the eigenvalues and eigenfunctions to identify the transition energies and estimate the probabilities of these transitions. The second Hamiltonian includes hyperfine splitting and interaction with an external magnetic field for a P1 center. The energy states of NV center and P1 center are written in the uncoupled bases $\vert m_S,m_I\rangle$.


The Hamiltonian for the NV center with the hyperfine interaction in an external magnetic field $\bf B$ directed along the $z$-axis, which coincides with the NV-axis, is \cite{Wood1}:
\begin{equation}
\hat H_\text{NV}=D \hat S_z^2  + \gamma_e \mathbf{B} \cdot \mathbf{\hat S} + \mathbf{\hat S} \cdot \hat{A}_{\text{NV}} \cdot \mathbf{\hat I} + Q_{\text{NV}}\hat I_z^2 - \gamma_{^{14}\text{N}} \mathbf{B} \cdot  \mathbf{\hat I},
 \label{eq1}
\end{equation}
where $\hat{A}_{\text{NV}}$ is a diagonal tensor of the hyperfine interaction between the electron spin $\bf S$ and the nuclear spin $\bf I$ in NV center,
\begin{equation}
\hat{A}_{\text{NV}} =
\left(
\begin{array}{ccc}
A_{\text{NV}\perp} & 0 & 0  \\ 
0 & A_{\text{NV}\perp} & 0 \\
0 & 0 & A_{\text{NV}\parallel}  
\end{array} 
\right) ,
\label{eqA}
\end{equation}
and $D = 2870$~MHz is the zero-field splitting of the ground-state electronic components with $m_S=0$ and $m_S=\pm 1$. The strengths of the electron and the nuclear spin interaction with the external magnetic field are determined by the gyromagnetic ratios $\gamma_e = 28.025$~MHz/mT and $\gamma_{^{14}\text{N}} = 3.077$~kHz/mT~\cite{Doherty2012}. The hyperfine magnetic dipole interaction parameters are $A_{\text{NV}\parallel} = - 2.14$~MHz, $A_{\text{NV}\perp} = -2.70$~MHz and the hyperfine electric quadrupole interaction parameter $Q_{\text{NV}} = - 4.96$~MHz \cite{Wood1}.

The Hamiltonian for the substitutional nitrogen (P1) is \cite{Wood1}:
\begin{equation}
\hat H_\text{P1}= \gamma_e \mathbf{B} \cdot \mathbf{\hat S}  - \gamma_{^{14}\text{N}} \mathbf{B} \cdot  \mathbf{\hat I} +  \mathbf{\hat S} \cdot \hat{A}_{^{14}\text{N}} \cdot \mathbf{\hat I} + Q_{^{14}\text{N}} \hat I_z^2 ,
 \label{eq2}
\end{equation}
where $\hat{A}_{^{14}\text{N}}$ is the diagonal hyperfine interaction tensor between the electron spin $\bf S$ and the nuclear spin $\bf I$ of $^{14}{\text{N}}$. $\hat{A}_{^{14}\text{N}}$ has the same mathematical structure as $\hat{A}_{\text{NV}}$ \eqref{eqA}. The Hamiltonian can be used for on-axis P1 centers as well as for off-axis P1 centers 
allowing for transverse magnetic field components $B_x$ and $B_y$.
The hyperfine interaction parameters for this Hamiltonian \eqref{eq2} are $A_{^{14}\text{N}\parallel} = 113.98$~MHz, $A_{^{14}\text{N}\perp} = 81.34$~MHz, and the quadrupole interaction parameter $Q_{^{14}\text{N}} = -3.97$~MHz \cite{Wood1}. From the hyperfine interaction's tensor-component values we see that the magnetic dipole part of the hyperfine interaction for the substitutional nitrogen is much stronger than that for the NV center. This can be explained by the fact that due to the spherical symmetry of the P1 center, the electrons are located close to the nitrogen atom. For the NV center, the electrons are mostly located around the vacancy, so they are separated by one lattice constant from the nitrogen atom~\cite{Doherty1}.

With these Hamiltonians one can calculate at what magnetic field strength a particular hyperfine energy-level splitting of the P1 center coincides with a particular hyperfine energy-level splitting of the NV center, which is the condition for cross-relaxation to occur.

\section{Experimental setup}

\begin{figure}[tb]
  \begin{center}
    \includegraphics[width=0.45\textwidth]{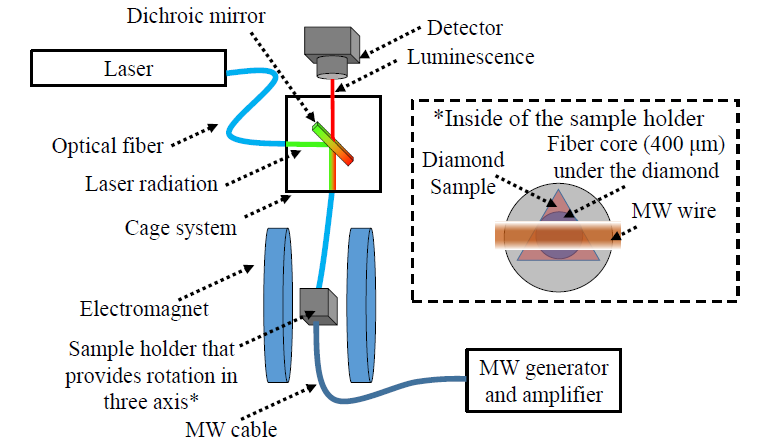}
  \end{center}
  \caption{Experimental setup.}
  \label{setup}
\end{figure}

The sample used in this experiment was a high-pressure, high-temperature (HPHT) diamond crystal with a (100) surface cut. To create NV centers it was irradiated with electrons. The irradiation dose was $10^{18}$ $e^-$ (cm$^{-2}$) at an irradiation energy of 10 MeV. After exposure the crystal was annealed for 3 hours at 750 $^{\circ}$C. The diamond crystal had a relatively high initial concentration of nitrogen $^{14}$N of around 200~ppm~\cite{Zheng1}.

Figure~\ref{setup} shows the experimental setup used, which is similar to the setup used in one of our previous experiments (see~\cite{Auzinsh1}). Light with a wavelength of 532 nm (Coherent Verdi Nd:YAG) was delivered to the sample via an optical fiber with a core diameter of 400 $\mu$m (numerical aperture 0.39). The same fiber was used to collect the red fluorescence light, which was separated from the residual green reflections with a dichroic mirror (Thorlabs DMLP567R) and a long-pass filter (Thorlabs FEL0600). The sensing volume for the experiment was about $0.4\times0.4\times0.35\text{~mm}^3$ determined by the dimensions of the optical fiber and the diamond. The sample was glued to the end of the optical fiber using an adhesive (Electron Microscopy Sciences CRYSTALBOND 509). The signals were recorded and averaged on a digital oscilloscope (Yokogawa DL6154).

To generate the low-frequency MW field (1.3~GHz to 1.6~GHz, $m_S = 0\longrightarrow m_S=-1$ transition) a function generator (SRS SG386) with power amplifiers (Minicircuits ZHL-2-12 and ZHL-16W-43+) for different MW frequency ranges were used. For the high-frequency MW field (4.1~GHz to 4.6 GHz, $m_S = 0\longrightarrow m_S=+1$ transition) the same function generator (SRS SG386) was combined with a different power amplifier (Minicircuits ZVE-3W-83+). The MW field was delivered using a copper wire in close proximity (at a distance of less than 1~mm) to the diamond sample, behind the diamond and the optical fiber.

\section{Signal analysis and results}

\subsection{Parameters used to characterize the ODMR signals}
\label{label1}

To align the diamond crystal in the external magnetic field, we initially measured the dependence of the laser induced fluorescence on the strength of the applied magnetic field, similarly to the previous work~\cite{Zheng1}. Aligning was done using the fluorescence features around 102.4~mT where the ground-state level anti-crossing (GSLAC) occurs. 

\begin{figure}[h!]
  \begin{center}
    \includegraphics[width=0.45\textwidth]{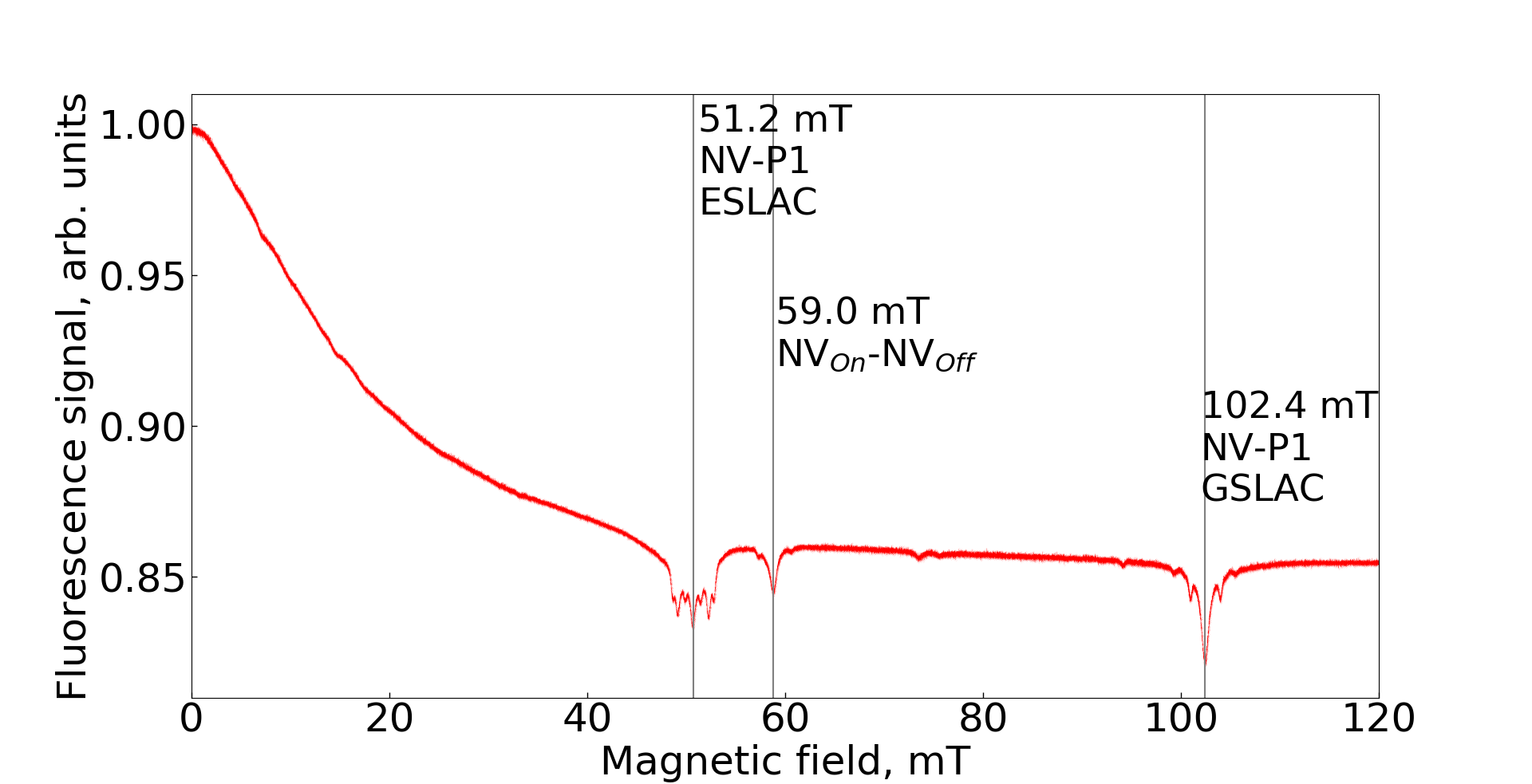}
  \end{center}
  \caption{NV Fluorescence signal dependence on the magnetic field strength.}
  \label{Fluorescence}
\end{figure}

A typical observed fluorescence signal as a function of the external magnetic field is depicted in FIG.~\ref{Fluorescence}. Narrow peaks or collections of peaks indicate the existence of cross-relaxation. Near 51.2~mT, cross-relaxation between the NV and P1 centers is observed.



At a magnetic field of around 59.0~mT cross-relaxation between the on-axis and the off-axis NV centers is observed. The off-axis directions are along the three possible axes in the diamond crystal that are not aligned with the applied external magnetic field. But these cross-relaxation processes were not studied in detail here.

\begin{figure*}[h!]
  \begin{center}
    \includegraphics[width=0.49\textwidth]{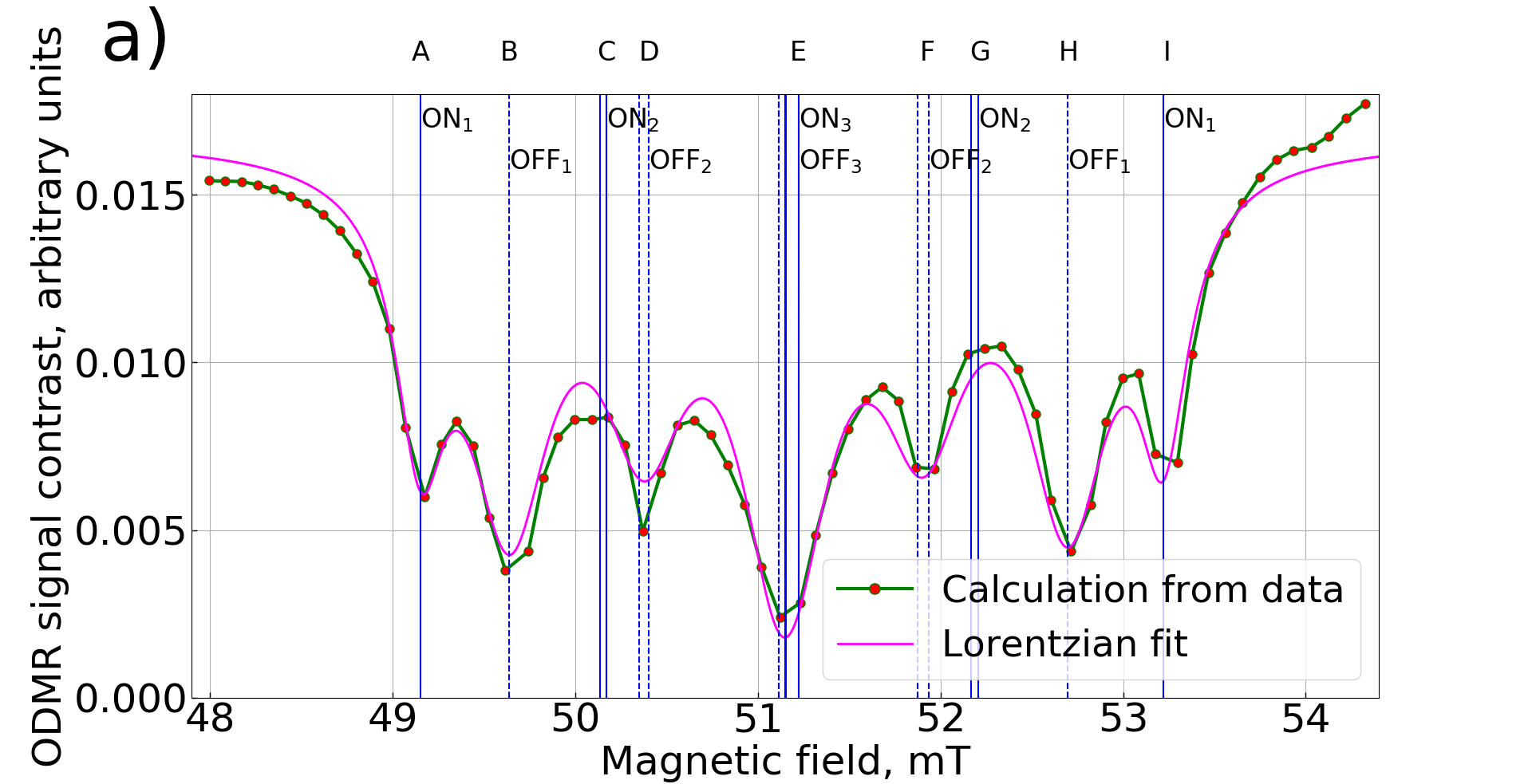}
    \hspace{0.1cm}
    \includegraphics[width=0.49\textwidth]{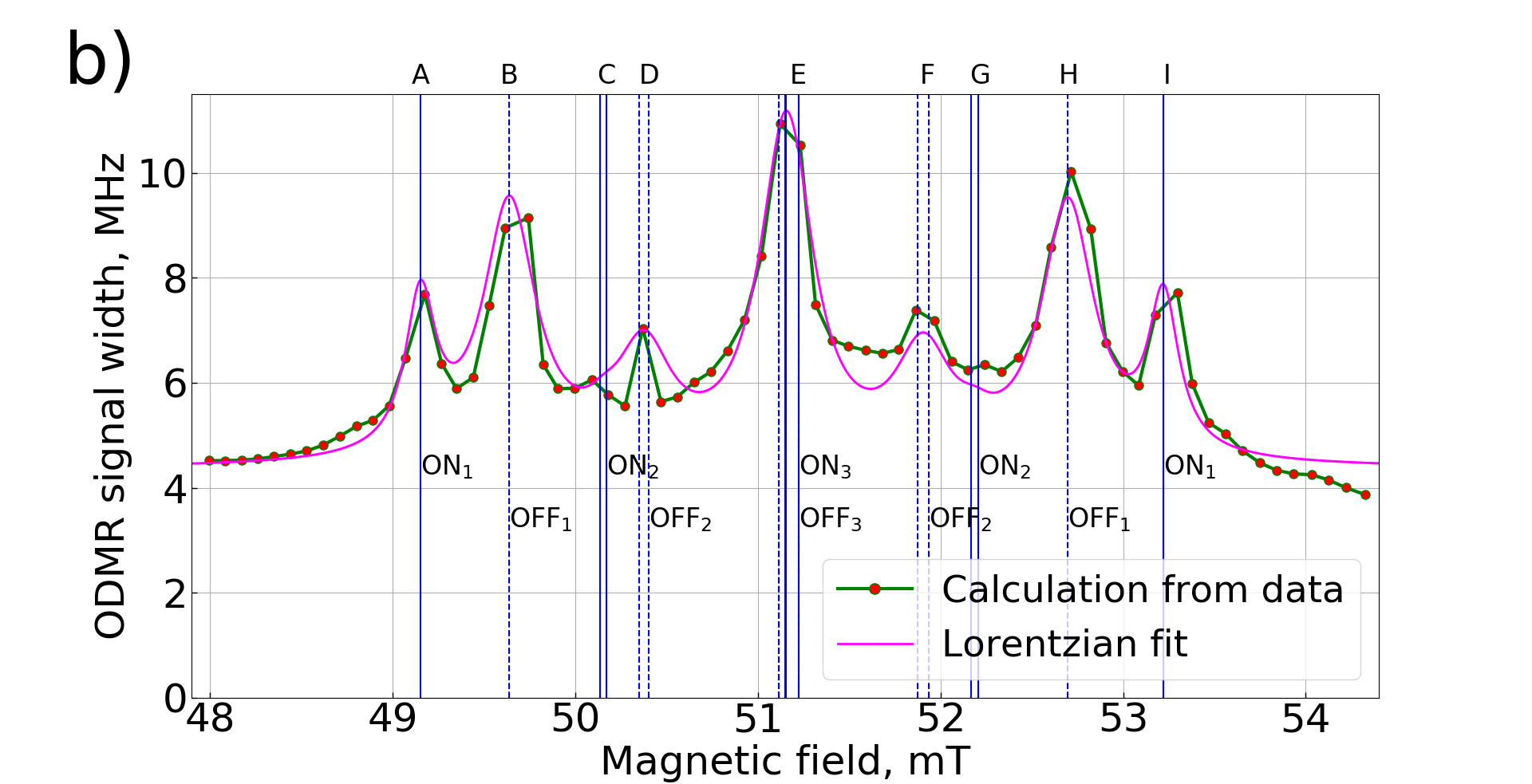}
  \end{center}
  \caption{NV center transitions $\vert m_S = 0, m_I\rangle\longrightarrow\vert m_S = -1, m_I\rangle$, the solid vertical lines indicate cross-relaxation resonance positions between the on-axis P1 centers and NV centers, the dotted vertical lines indicate cross-relaxation resonance positions between off-axis P1 centers and NV centers. Images a) and b) show the measured ODMR signal resonance contrast and width (respectively) that are extracted from the experimental signals.}
  \label{minus1}
\end{figure*}
\begin{figure*}[h!]
  \begin{center}
    \includegraphics[width=0.49\textwidth]{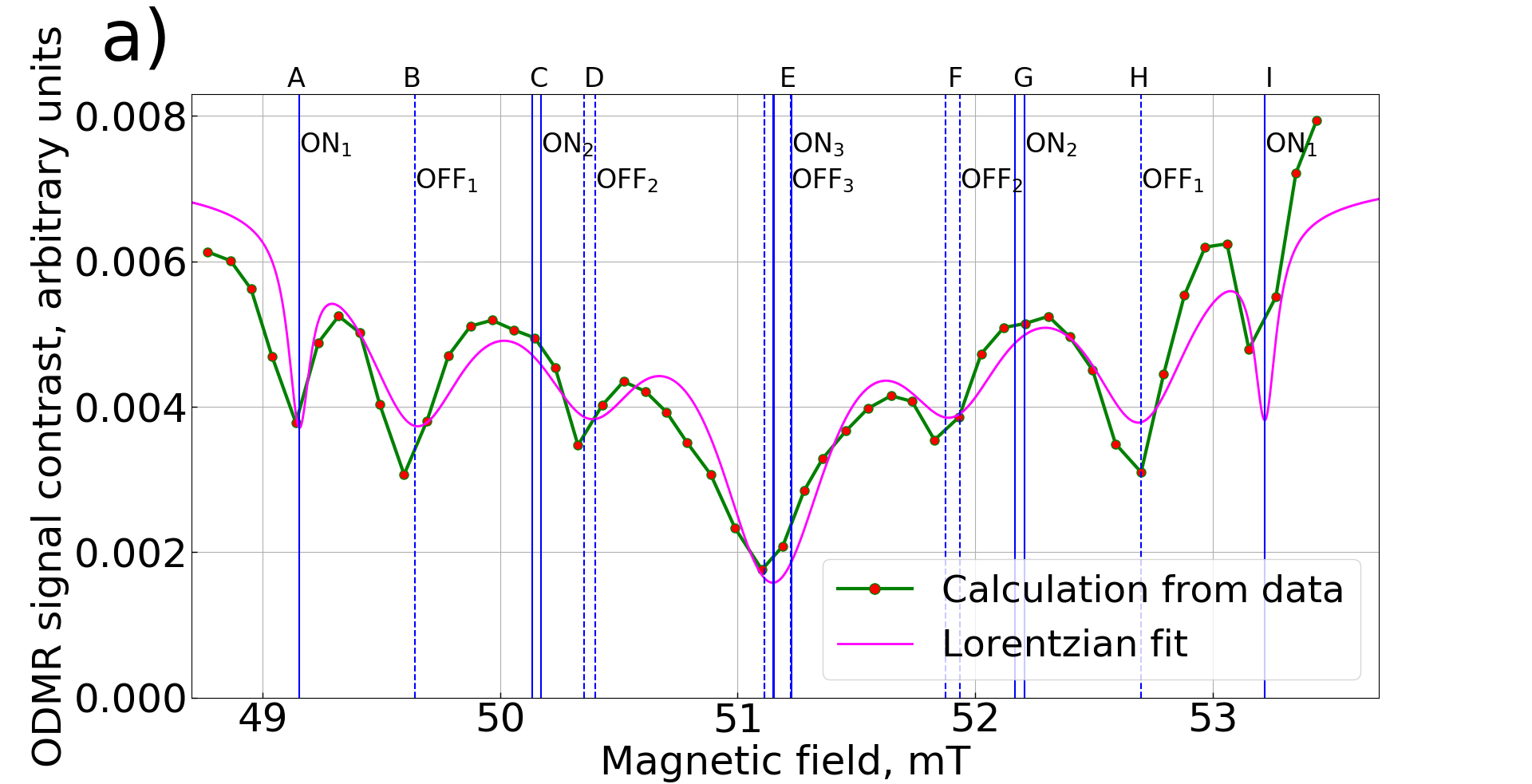}
    \hspace{0.1cm}
    \includegraphics[width=0.49\textwidth]{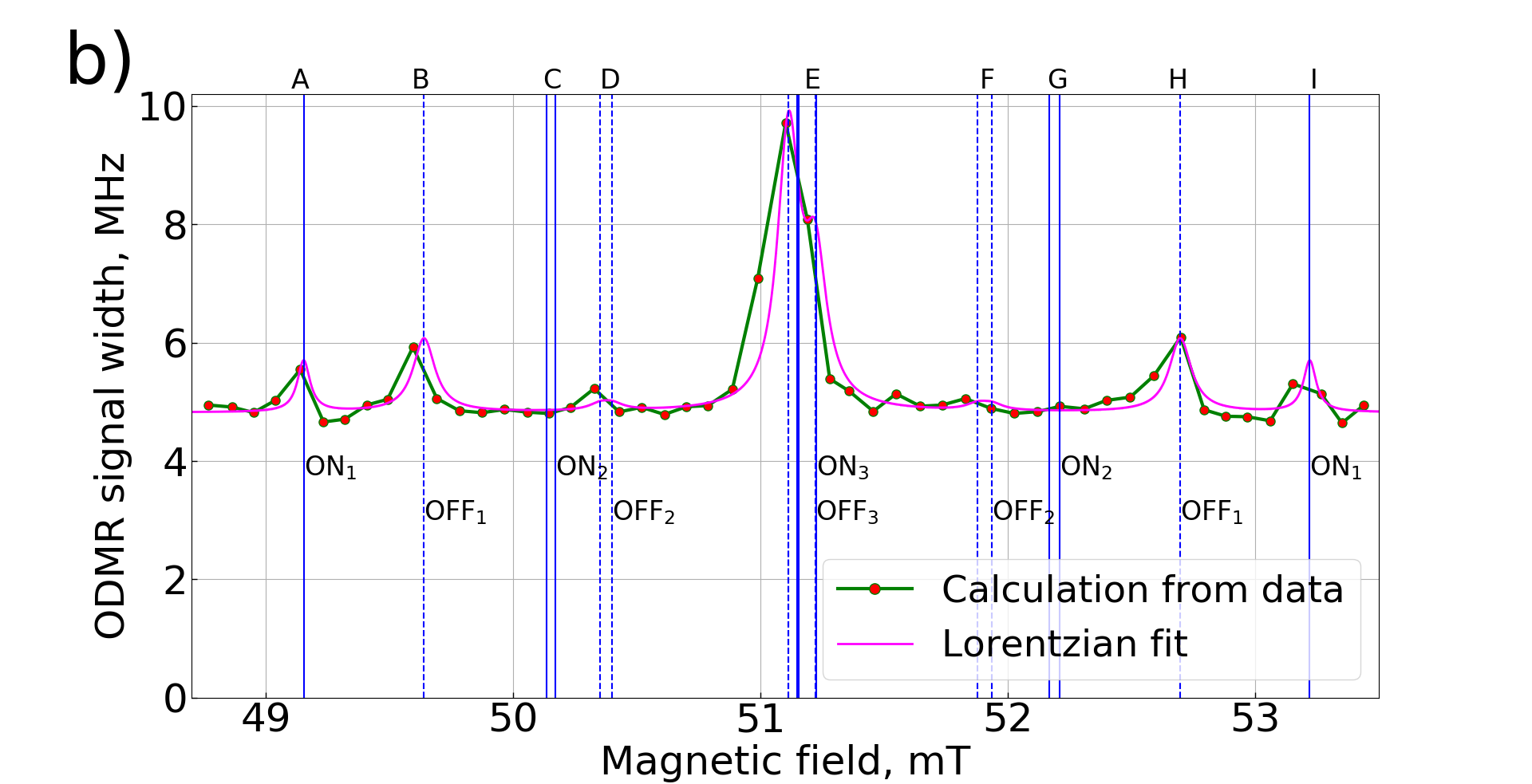}
  \end{center}
  \caption{NV center transitions $\vert m_S = 0, m_I\rangle\longrightarrow\vert m_S = +1, m_I\rangle$, the solid vertical lines indicate cross-relaxation resonance positions between the on-axis P1 centers and NV centers, the dotted vertical lines indicate cross-relaxation resonance positions between off-axis P1 centers and NV centers. Images a) and b) show the measured ODMR signal resonance contrast and width (respectively) that are extracted from the experimental signals.}
  \label{plus1}
\end{figure*}
\begin{figure}[h!]
  \begin{center}
    \includegraphics[width=0.45\textwidth]{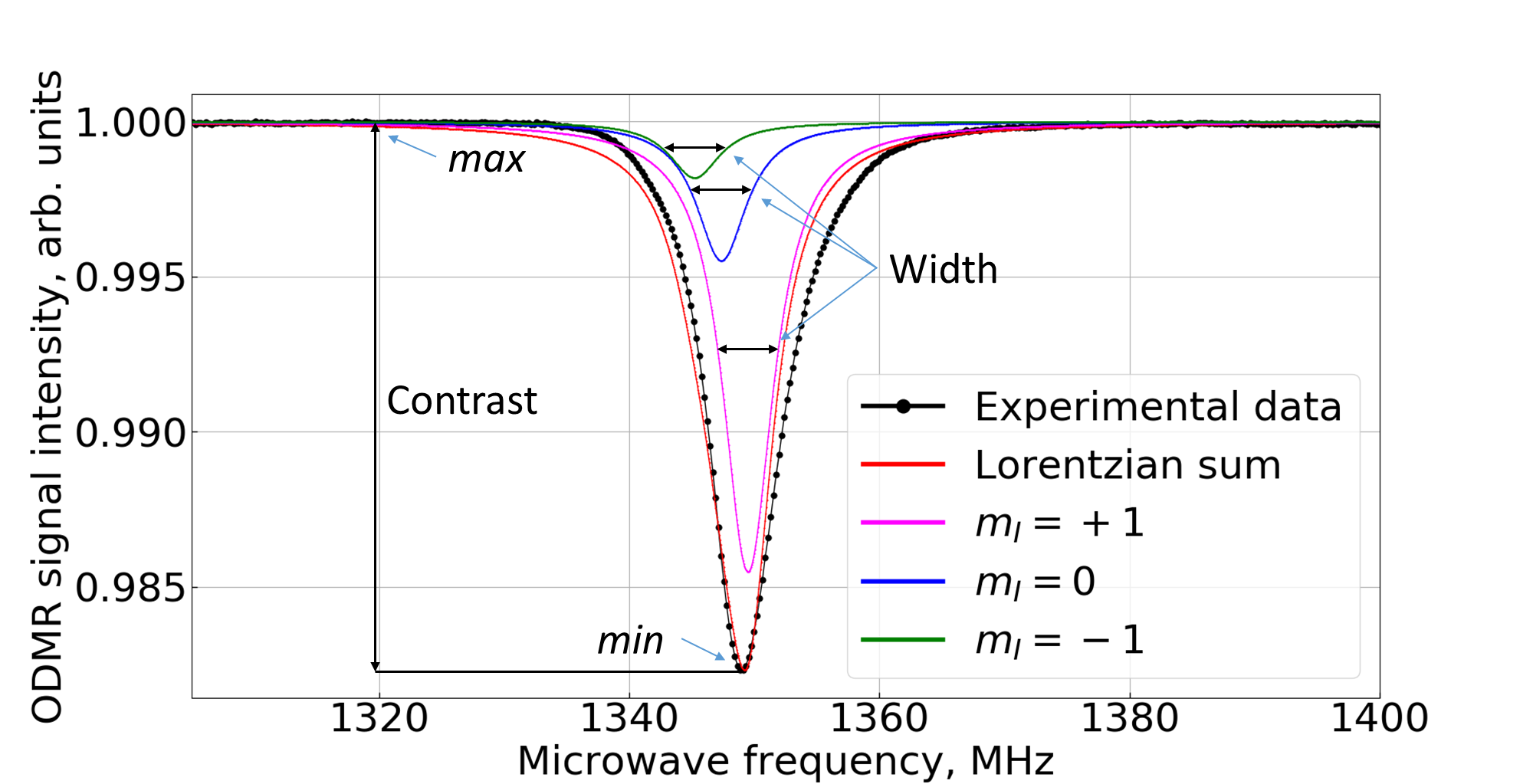}
  \end{center}
  \caption{Typical, measured ODMR signal (normalized, arbitrary units) from the NV centers at around 51.2~mT external magnetic field for the $m_S = 0 \longrightarrow m_S = -1$ transition with unresolved hyperfine structure, showing the full width at half maximum for the three Lorentzians and the contrast of the ODMR signal.}
  \label{ODMR}
\end{figure}

We measured ODMR signals and from these signals obtained the dependence of the signal-contrast and width on the external magnetic field strength. 
The ODMR signal contrast (FIG.~\ref{minus1}a and FIG.~\ref{plus1}a) was calculated as $1-\frac{min}{max}$ 
(see FIG.~\ref{ODMR}).
We measured the position, contrast, and width of cross-relaxation signals for two sets of transitions.

Initially we used ODMR signals in the MW frequency range of $1.10 - 1.40$~GHz where the NV center's electron spin transitions between the ground state magnetic sublevels $\vert m_S = 0, m_I\rangle$ and $\vert m_S = -1, m_I\rangle$ are involved, meaning that both the $\vert m_S = 0, m_I\rangle$ and the $\vert m_S = -1, m_I\rangle$ NV center's sublevels of the ground state MW transition are involved in the cross-relaxation process with the P1 center's $\vert m_S=-\frac{1}{2},m_I\rangle\iff\vert m_S=\frac{1}{2},m_I\rangle$ transition.

Secondly, we registered the ODMR signals in the MW frequency range of $4.10 - 4.40$~GHz, which corresponds to the electron spin transitions between the magnetic sublevels $\vert m_S = 0, m_I\rangle$ and $\vert m_S = +1, m_I\rangle$ in the NV center. In this transition only the lower ground state level of the NV center $\vert m_S = 0, m_I\rangle$ is directly involved in the cross-relaxation process with the P1 center's $\vert m_S=-\frac{1}{2},m_I\rangle\longrightarrow\vert m_S=\frac{1}{2},m_I\rangle$ transition, while the higher $\vert m_S=+1, m_I\rangle$ ground state energy level of the NV center is unperturbed.

We observed two peculiarities in FIG.~\ref{minus1} and FIG.~\ref{plus1}. First, for the NV transition $\vert m_S=0, m_I\rangle\longrightarrow\vert m_S=-1, m_I\rangle$ the average ODMR signal contrast is at least two times larger than that for the $\vert m_S=0, m_I\rangle\longrightarrow\vert m_S=+1, m_I\rangle$ transition. Second, if  we compare changes in the ODMR signal contrast and width for a specific transition, we see that the central peak at 51.2~mT (E in FIG.~\ref{minus1} and FIG.~\ref{plus1}) exhibits the largest changes in contrast and width with the magnetic field. 


The first peculiarity can be attributed to the above mentioned fact that for $\vert m_S=0, m_I\rangle\longrightarrow\vert m_S=-1, m_I\rangle$ transition in cross-relaxation conditions the relaxation rate for both the initial and the final states increases, but for $\vert m_S=0, m_I\rangle\longrightarrow\vert m_S=+1, m_I\rangle$ transition only the initial level is broadened by the increased relaxation rate.

The second peculiarity is due to the fact that, as will be shown in more detail in the next Section, this central peak corresponds not to one, but to several transitions in P1 center, see FIG.~\ref{peaks}.


The excited-state energy levels undergo crossing at around 51.2~mT as well, so the transition probabilities from the excited state to the ground state change around these magnetic field values due to the crossing and mixing of the excited state $\vert m_S=-1\rangle$ and $\vert m_S=0\rangle$ energy levels. This crossing might have an effect on the nuclear spin polarization of the NV center. We assume that this effect is not dominant in this case since the ESLAC occurs over a relatively broad frequency range that extends across the entire range of magnetic field values used in this experiment~\cite{Doherty1, Wrachtrup1}. Also, the nuclear spin polarization of the NV center remains high at the magnetic field values around 51.2~mT with only a small decrease at the ESLAC point~\cite{Bud3}. Furthermore, in samples with low P1 concentration, the fluorescence structure at the magnetic field that leads to ESLAC consists mainly of one peak, which indicates that the rich structure originates mostly from cross-relaxation processes~\cite{Zheng1}.

A detailed analysis of the cross-relaxation transitions involved in the formation of the peaks detected in the signals are presented in FIG.~\ref{peaks} and reflected in the analysis of this figure. 



Cross-relaxation transitions that involve individual hyperfine structure components of the NV centers are hidden under the unresolved ODMR signal measured in the vicinity of the magnetic field of $51.2$~mT (see FIG.~\ref{ODMR}). 


Similar cross-relaxation transitions have been studied before in~\cite{Bud2, Wood1} using a substantially different method. The authors measured the dependence of the spin-relaxation time $T_1$ on the magnetic field strength in the vicinity of the cross-relaxation resonances by means of sequences of optical and MW pulses. Furthermore, in their analysis the authors of~\cite{Bud2, Wood1} neglected the hyperfine structure in NV centers which, as will be shown below, can lead to an interpretation that does not describe all the features of the observed cross relaxation signals.

At the start of the signal analysis it is worth noting that in the results of our experiment and the experiments of~\cite{Bud2, Wood1} there are some qualitative differences. Peaks denoted as D and F in our study (see FIG.~\ref{minus1} and FIG.~\ref{plus1}), are associated with transitions that in the supplementary information of~\cite{Bud2} are called  disallowed transitions. These transitions, according to~\cite{Bud2}, correspond to simultaneous flip-flop between the on-axis P1 electron and nuclear spins. In~\cite{Bud2}, these peaks are noticeably weaker than in this study.

When we take into account the nuclear spin of the NV center as well, a new partner appears in the analysis of the angular momentum projection balance. As a result, we will not be using the notion of allowed and disallowed transitions in the present analysis.

\subsection{Simultaneous interaction of MW with NV and P1 centers}
\label{jaanjanodalja}
We think that the difference in the relative intensity of the peaks is due to the fact that at the magnetic field strength where cross-relaxation occurs and the energy-level splittings in P1 and NV centers are equal, the MW field that is applied to measure ODMR signals is resonantly coupled not only to the hyperfine transitions of the NV center, but of the $\vert m_S=0, m_I\rangle\longrightarrow\vert m_S=-1, m_I\rangle$ MW hyperfine transition of the P1 center as well.




Let us make rough estimates on how MW radiation simultaneously coupled to the resonant transitions for both partners of the cross-relaxation process can influence the signals.

\begin{figure}
    \centering
    \includegraphics[width=0.7\columnwidth]{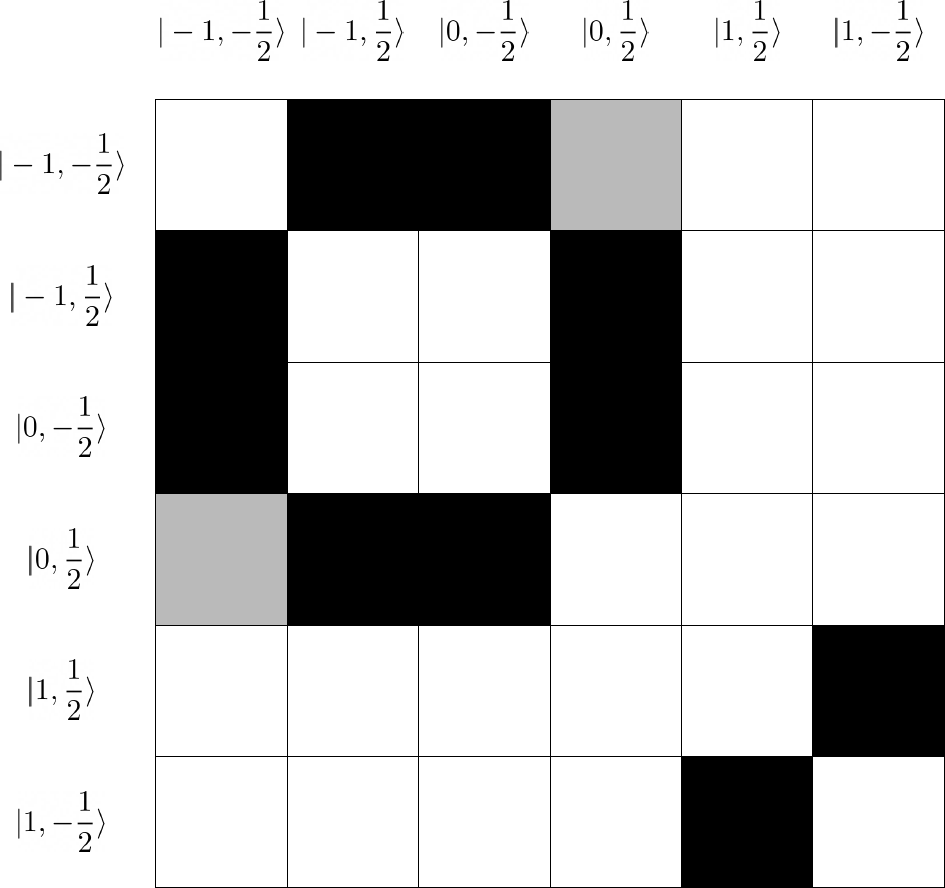}
    \caption{Rotating-frame Hamiltonian after fast-oscillating terms have been set to zero for the energy matched (only the black squares) and unmatched (black and grey squares indicating the matrix elements related to the MW) cases. The kets surrounding the figure describe the ordering of the $\vert m_\text{NV}, m_\text{P1}\rangle$ electron-spin basis states.}
    \label{fig:matarr}
\end{figure}

To investigate the impact of the two different approaches (pulsed MW vs CW ODMR experiment) we developed a model  that includes a P1 center dipole-coupled to a proximal NV center (nuclear contributions are ignored). The Hamiltonian that describes the evolution of this system consists of three parts:
\begin{equation}
    \hat{H}_\text{tot} = \hat{H}_\text{NV} + \hat{H}_\text{P1} + \hat{H}_\text{int},
    \label{eq4}
\end{equation}
where $\hat{H}_\text{NV}$ and $\hat{H}_\text{P1}$ are the Hamiltonians of the NV and P1 centers, respectively, as defined in equations \eqref{eq1} and \eqref{eq2} without the nuclear contributions, and $\hat{H}_\text{int}$ describes the dipolar interactions between the two which can be written as:
\begin{equation}
\begin{aligned}
\hat{H}_\text{int}=\mathcal{K}_1\left[\frac{\mathbf{\hat{S}_\text{NV}}\cdot\mathbf{\hat{S}_\text{P1}}}{r^3}-3\frac{1}{r^5}\left(\mathbf{\hat{S}_\text{NV}}\mathbf{r}\right)\cdot\left(\mathbf{\hat{S}_\text{P1}}\mathbf{r}\right)\right],
\end{aligned}
\end{equation}
where $\mathcal{K}_1$ is a constant that describes the interaction strength, $\mathbf{r}$ is the vector that connects the two pointlike defects in diamond. If, in addition, the MW field is applied, its action is described as:
\begin{equation}
\begin{aligned}
\hat{H}_\text{CW} = &\hat{H}_\text{tot} +\\
&+\Omega_1 \cos\left(\omega t\right)\hat{S}_{\text{NV} x} + \Omega_2 \cos\left(\omega t\right)\hat{S}_{\text{P1} x},
\end{aligned}
\end{equation}
where $\omega$ is the applied MW frequency and $\Omega_{1,2}$ are the corresponding Rabi frequencies. For this model calculation we can assume $\Omega_{1,2}$ to be equal, as the $g$-factors of P1 and NV centers are similar, and $\omega$ is assumed to be resonant with the corresponding spin transitions in both NV and P1 centers. To simplify the problem it is treated in the rotating frame \cite{hartmann_nuclear_1962,rovnyak_tutorial_2008},
where the effective Hamiltonian can be obtained by the following transformation:
\begin{align}
\begin{split}
\hat{H}_\text{rot} = &e^{i(\hat{H}_\text{P1}+\hat{H}_\text{NV})t}\hat{H}_\text{CW}e^{-i(\hat{H}_\text{P1}+\hat{H}_\text{NV})t}-\\
&-(\hat{H}_\text{P1}+\hat{H}_\text{NV}).
\label{eq7}
\end{split}
\end{align}
A visual representation of the obtained interaction matrices in the $\vert m_\text{NV}, m_\text{P1} \rangle$ electron spin basis are shown in FIG.~\ref{fig:matarr}.

\begin{figure}
    \centering
    \includegraphics[width=\columnwidth]{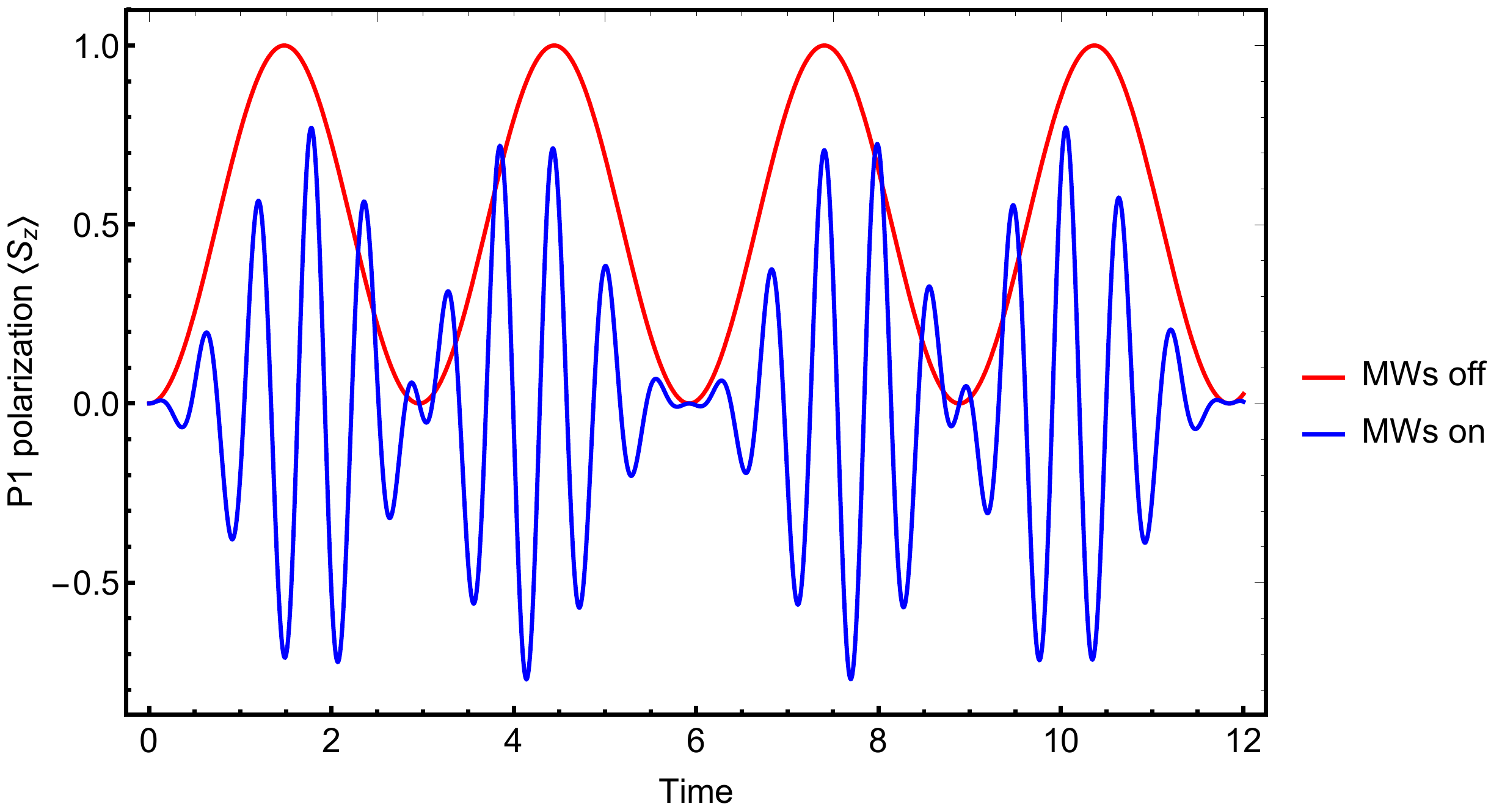}
    \caption{Expectation value $\langle\hat{S}_{\text{P1}z}\rangle$ as a function of time with and without a microwave field.}
    \label{fig:polar_dep}
\end{figure}

Initially the system is assumed to be totally polarized in the electron spin $\vert m_\text{NV} = 0 \rangle$ state while the P1 spin is assumed to be unpolarized. Then the system is allowed to evolve under the influence of the effective rotating frame Hamiltonian and the P1 polarization is monitored (defined as the expectation value of $\langle \hat{S}_{\text{P1}z}\rangle$). Two cases, one with no MW field and one with a field much stronger than the average NV-P1 dipole coupling strength ($\mathcal{K}_1$) value, are shown in FIG.~\ref{fig:polar_dep}. While the peak polarization of the P1 centers is slightly higher without the MW field, the transfer rate due to Rabi oscillations is much more rapid with the field on.



We can assume that the rapid changes in the P1 electron spin polarization when the MW field is on can be interpreted as an increase of  the spin relaxation rate in the presence of the MW field. This increased effective relaxation decreases the ODMR signal contrast and increases the signal width.


As far as the MW signal influences both partners in the cross-relaxation process we cannot expect that the relative peak intensities in pulsed experiments~\cite{Bud2, Wood1} and this ODMR experiment will be the same.
The reason is the following: in a pulsed experiment, after the initial polarization of the NV centers, they are left to relax during a time of field-free interaction with the P1 centers. And after a certain interaction time the laser field is switched on for readout. On the contrary, in our experiment, the optical and the MW fields are on all of the time.
And finally, most probably the introduction of the neglected nuclear spin, which is not directly coupled to the MW radiation into the model, will not qualitatively alter the conclusions derived from this simple model. 

Let us now come back to the analysis of the influence of the NV nuclear spin on the cross-relaxation signals.
It is seen experimentally and understood in the model simulations that, due to hyperfine mixing in the external magnetic field, most of the population of the NV center is polarized to the $\vert m_S=0, m_I=+1\rangle$ state, which does not mix with other states~\cite{Bud3,Ivady}.



At the temperatures used in the experiment the P1 energy levels are populated almost equally. The relative difference of the populations is around $2\cdot 10^{-4}$, determined by the Boltzmann distribution, with the $\vert m_S=-\frac{1}{2}\rangle$ energy level being slightly more populated (due to the fact that it has a lower energy in the applied external magnetic field).




\subsection{Detailed analysis of the different cross--relaxation transitions}

The ODMR signal contrasts depicted in FIG.~\ref{minus1}a and FIG.~\ref{plus1}a 
are obtained from the ground-state unresolved hyperfine MW transitions (see FIG.~\ref{ODMR}). The magnetic field value at each point is determined from the ODMR peak position (the mean position value of the three hyperfine components that make up the unresolved ODMR spectral peak).
The full width at half maximum of the individual assumed Lorentzian-shaped components was determined by fitting the data with a sum of three Lorentzian functions with equal widths.
The distance between them (the difference in energy between the hyperfine components) was calculated~\cite{Auzinsh1} using equations (4). Figures~\ref{minus1}b and ~\ref{plus1}b show the width of the individual Lorentzians (FIG.~\ref{ODMR}) as the width of the resonances changes when at certain magnetic field values cross-relaxation conditions (equal energy splitting) are met.

The magnetic field step in the plots of ODMR signal contrast and width was constrained by our experimental setup. The smallest step was about 0.09~mT (the resolution of the power supply for the electromagnet).

\begin{figure*}[h!]
  \begin{center}
    \includegraphics[width=0.99\textwidth]{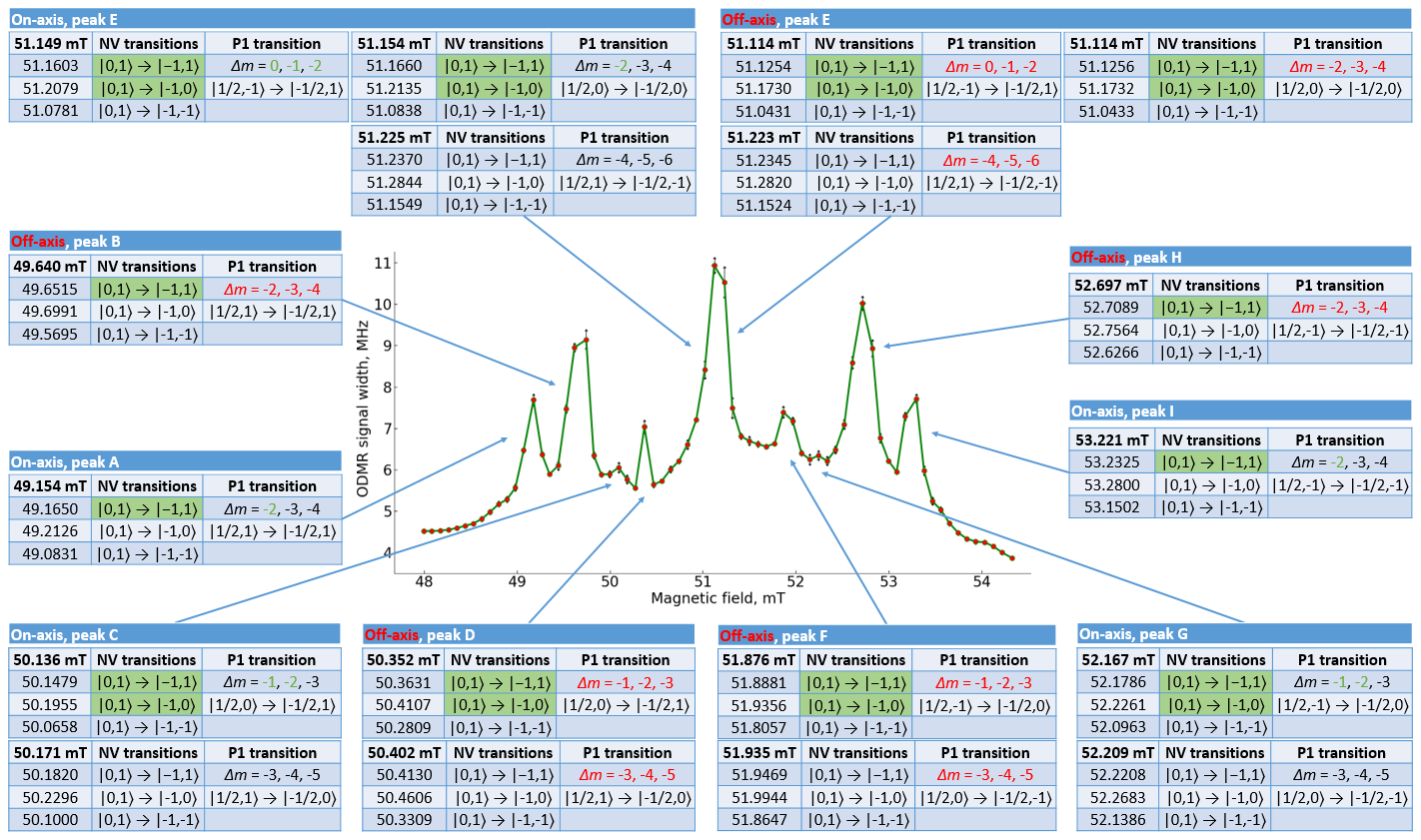}
  \end{center}
  \caption{ODMR signal width as a function of the applied external magnetic field. The red points are calculated from the experimental data for the $\vert m_S = 0, m_I\rangle\longrightarrow\vert m_S = -1, m_I\rangle$ NV center transitions. The tables around the graph show the mean magnetic field values (bold) of three possible transitions for an NV center in combination with three transitions for the P1 center (the same transition at slightly different magnetic field values). The NV center and P1 center transitions are written in the $\vert m_S, m_I\rangle$ bases. Cells and numbers marked in green represent transitions noted to have the total spin projection (magnetic quantum number) change's absolute value $\vert\Delta m\vert = \vert\Delta m_{S_\text{NV}} + \Delta m_{I_\text{NV}} + \Delta m_{S_\text{P1}} + \Delta m_{I_\text{P1}}\vert$  less than or equal to 2 (this is true only when the projections are along the same axis, otherwise this can be taken only as a classification notation). The off-axis transitions are marked red in the tables.}
  \label{peaks}
\end{figure*}

The marked positions in FIG.~\ref{minus1} and FIG.~\ref{plus1} correspond to the transitions from the most populated hyperfine structure state $\vert m_S=0,m_I=+1\rangle$ \cite{Bud3} to some other empty or much less populated states (see FIG.~\ref{peaks}).

The solid, smooth curve (magenta) in FIG.~\ref{minus1} and FIG.~\ref{plus1} shows the result of a Lorentzian fit of the contrast and width data for these cross-relaxation peaks. Each Lorentzian peak's maximum position of the fit curve was assumed to correspond to the average of the calculated magnetic field values for the $\vert m_S=0, m_I=+1\rangle$ to the $\vert m_S=-1, m_I=0, \pm 1\rangle$ NV center transitions (bold magnetic field values in the in-figure boxes in FIG.~\ref{peaks}).


Five large-amplitude cross-relaxation peaks (peaks A, B, E, H, I in figures~\ref{minus1} and \ref{plus1}) and four smaller peaks (C, D, F, G in figures~\ref{minus1} and \ref{plus1}) are observed.

The analysis of the cross-relaxation peaks in terms of angular momentum projection conservation is complicated by the fact that for off-axis P1 centers the natural direction of the quantization axis is the direction from the P1 center to the NV center with which the cross-relaxation energy exchange is expected -- the P1-NV direction.
The spin projection magnetic quantum numbers $m_S$ and $m_I$ defined in the P1-NV reference system are not good quantum numbers any more with respect to the NV center's own reference system, which is the NV center's axis direction and coincides in our experiment with the direction of external magnetic field $\bf B$. So the angular momentum quantum projection numbers used in the text should be considered rather as labels to indicate states. 


The group of four overall larger peaks that are symmetric to the central one correspond to on-axis P1 cross-relaxation transitions (peaks A and I) and off-axis P1 cross-relaxation transitions (peaks B and H) that conserve nuclear spin projection in P1 during the cross-relaxation processes (see FIG.~\ref{peaks}). Peaks B and H in this group of four peaks have a larger amplitude than peaks A and I. This amplitude difference is plausible because there is a three times higher probability to have an off-axis P1 center in the vicinity of a particular NV center than to have an on-axis P1 center.

Two smallest peaks -- C and G correspond to P1 on-axis cross-relaxation transitions in which the nuclear spin projection of the P1 center is not conserved. Slightly more intense peaks correspond to similar cross-relaxation transitions with a change in nuclear spin in the off-axis P1 centers, which are three times more probable.

Looking at the balance of the angular momentum projections including the electron spin as well as the nuclear spin of both partners -- the NV center and the P1 center, it is seen from the legend in FIG.~\ref{peaks}  that this projection is not conserved. There are several reasons why spin projection nonconserving transitions in cross-relaxation are possible. Fist, the phonons and the crystal lattice itself can take or provide some of the angular momentum~\cite{Bloembergen1}. Second, our experiment involves MW photons, which can provide angular momentum.

Coming back to the cross-relaxation signals, as was pointed out in~\cite{Bud2, Wood1}, an intense peak appears around the magnetic field strength values of 51.2~mT -- the central peak E (see FIG.~\ref{minus1} and FIG.~\ref{plus1}). This peak corresponds simultaneously to a number of on-axis and off-axis transitions (FIG.~\ref{peaks}), which accounts for its rather large amplitude.

\section{Conclusions}

In this study we have demonstrated that ODMR measurements on NV centers in diamond crystals can be used successfully to study cross-relaxation processes between NV centers and P1 centers.
To interpret the signals produced by these resonances we accounted for the hyperfine structure in both interaction partners, NV centers and P1 centers. In addition, we accounted for the nuclear spin polarization in the NV centers, which occurs during the process of optical polarization of the electron spin of the NV center in the magnetic field due to strong hyperfine level mixing. Using these tools we were able to analyze in detail the cross-relaxation resonance peak positions.


Our measurements confirm the presence of five main cross-relaxation transitions (see peaks A, B, E, H, I in figures~\ref{minus1} and  \ref{plus1}). This result is in a qualitative agreement with previous studies~\cite{Wood1, Laraoui1, Lange1, Knowles1}. However, in contrast to these cited studies, in our experiment the probabilities of transitions that in~\cite{Bud2} are classified as ``disallowed'' (transitions C, D, F, and G) are observed as being rather high. The reason for this difference is that our experiments were performed under continuous MW radiation, whose absorbed photons continuously provide angular momentum to enable these otherwise ``disallowed'' transitions.




Cross-relaxation occurs when the energy splitting between a pair of hyperfine magnetic sublevels in an NV center coincides with a hyperfine magnetic sublevel splitting in substitutional nitrogen. Under these conditions the relaxation rate in the NV center effectively increases, ODMR signal contrast decreases, and resonances become broadened.


This interpretation of the signal origin is supported by the experimental evidence. When ODMR signals are observed between the hyperfine levels of the electron spin transition $\vert m_S = 0\rangle\longrightarrow\vert m_S = -1\rangle$ (see FIG. \ref{minus1}), the change in the signal amplitude at the resonance point is around two times larger than when the resonance is observed on the $\vert m_S = 0\rangle \longrightarrow \vert m_S = +1\rangle$ transition (see FIG.~\ref{plus1}). As can be seen from FIG.~\ref{energy_levels}, when we observe the $\vert m_S = 0\rangle \longrightarrow \vert m_S = -1\rangle$ resonances, both levels of the transition in the NV center interact resonantly with the transition in the substitutional nitrogen. On the contrary, the $\vert m_S = 0\rangle\longrightarrow \vert m_S = +1\rangle$ ODMR transition does not interact resonantly with any particular transition in the substitutional nitrogen. Only the lower level $\vert m_S = 0\rangle$ of the NV transition at some particular magnetic field strength is influenced by the cross-relaxation process. As a result, in contrast to the former case, in this particular case, only one level of the transition is broadened due to cross-relaxation, and as a result the signal contrast is diminished less effectively and the observed signal broadening is weaker.

As shown in the model calculation above (Subsection~\ref{jaanjanodalja}), this difference in effective relaxation rates is further enhanced by the fact that in the case of $\vert m_S = 0\rangle\longrightarrow \vert m_S = -1\rangle$ resonances, the MW field under resonance conditions effectively interacts with both partners, the NV center and the substitutional nitrogen, leading to interaction dynamics (see FIG.~\ref{fig:polar_dep}) that can be understood as a further increase in the effective relaxation rate. This further enhancement is not present for the $\vert m_S = 0\rangle\longrightarrow\vert m_S = +1\rangle$ transition when the MW field is not in resonance with the transition in the P1 center.

The results presented in this paper serve to develop new ways of using NV centers in electron and nuclear magnetic resonance methods, as well as for studies of nuclear spin hyperpolarization with NV centers in diamond. Furthermore, these results can be important for cross-relaxation studies involving other point-like defects in a diamond crystal, optimization of experimental conditions for microwave-free magnetometry, and microwave-free nuclear-magnetic-resonance probes.



\section{Acknowledgements}
We gratefully acknowledge the financial support from the Base/Performance Funding Project Nr. ZD2010/AZ27, AAP2015/B013 of the University of Latvia. A.~Berzins acknowledges support from the PostDoc Latvia Project Nr. 1.1.1.2/VIAA/1/16/024 "Two-way research of thin-films and NV centres in diamond crystal".

\bibliography{Crossrelaxation}

\begin{thebibliography}{10}

\bibitem{Popkin1}
Gabriel Popkin.
\newblock {\em Science}, 354(6316):1091, 2016.

\bibitem{Ladd}
T.~D. Ladd, F.~Jelezko, R.~Laflamme, Y.~Nakamura, C.~Monroe, and J.~L. O'Brien.
\newblock Quantum computers.
\newblock {\em Nature}, 464(4 March):45--53, 2010.

\bibitem{Heshami1}
Khabat Heshami, Duncan~G. England, Peter~C. Humphreys, Philip~J. Bustard,
  Victor~M. Acosta, Joshua Nunn, and Benjamin~J. Sussman.
\newblock {\em Journal of Modern Optics}, 63(20):2005 -- 2028, 2016.

\bibitem{Zheng1}
Huijie Zheng, Georgios Chatzidrosos, Arne Wickenbrock, Lykourgos Bougas, Reinis
  Lazda, Andris Berzins, Florian~Helmuth Gahbauer, Marcis Auzinsh, Ruvin
  Ferber, and Dmitry Budker.
\newblock {\em Proc. SPIE 10119, Slow Light, Fast Light, and Opto-Atomic
  Precision Metrology X}, (101190X), 2017.

\bibitem{Taylor}
J.~M. Taylor, P.~Cappellard, L.~Childress, L.~Jiang, D.~Budker, P.~R.
  Hemmerand~A. Yacoby, R.~Walsworth, and M.~D. Lukin.
\newblock High--sensitivity diamond magnetometer with nanoscale resolution.
\newblock {\em Nature Physics}, 4(October):810--816, 2008.

\bibitem{Dolde}
F.~Dolde, H.~Fedder, M.~W. Doherty, T.~N{\"o}bauer, F.~Rempp,
  G.~Balasubramanian, T.~Wolf, F.~Reinhard, L.~C.~L. Hollenberg, F.~Jelezko,
  and J.~Wrachtrup.
\newblock Electric-field sensing using single diamond spins.
\newblock {\em Nature Physics}, 7(17 April):459--463, 2011.

\bibitem{Clevenson1}
Hannah Clevenson, Matthew~E. Trusheim, Carson Teale, Tim Schr{\"o}der, Danielle
  Braje, and Dirk Englund.
\newblock {\em Nature Physics}, 11:393 -- 397, 2015.

\bibitem{Acosta}
V.~M. Acosta, E.~Bauch, M.~P. Ledbetter, A.~Waxman, L.-S. Bouchard, and
  D.~Budker.
\newblock Temperature dependence of the nitrogen-vacancy magnetic resonance in
  diamond.
\newblock {\em Phys. Rev. Lett.}, 106:209901, 2011.

\bibitem{Plenio1}
P.~{Fern{\'a}ndez-Acebal}, O.~{Rosolio}, J.~{Scheuer}, C.~{M{\"u}ller},
  S.~{M{\"u}ller}, S.~{Schmitt}, L.~P. {McGuinness}, I.~{Schwarz}, Q.~{Chen},
  A.~{Retzker}, B.~{Naydenov}, F.~{Jelezko}, and M.~B. {Plenio}.
\newblock {Toward Hyperpolarization of Oil Molecules via Single Nitrogen
  Vacancy Centers in Diamond}.
\newblock {\em Nano Letters}, 18(3):1882--1887, Mar 2018.

\bibitem{Pines1}
Jonathan~P. {King}, Keunhong {Jeong}, Christophoros~C. {Vassiliou}, Chang~S.
  {Shin}, Ralph~H. {Page}, Claudia~E. {Avalos}, Hai-Jing {Wang}, and Alexander
  {Pines}.
\newblock {Room-temperature in situ nuclear spin hyperpolarization from
  optically pumped nitrogen vacancy centres in diamond}.
\newblock {\em Nature Communications}, 6:8965, Dec 2015.

\bibitem{Broadway1}
David~A. {Broadway}, Jean-Philippe {Tetienne}, Alastair {Stacey}, James D.~A.
  {Wood}, David~A. {Simpson}, Liam~T. {Hall}, and Lloyd C.~L. {Hollenberg}.
\newblock {Quantum probe hyperpolarisation of molecular nuclear spins}.
\newblock {\em Nature Communications}, 9:1246, Mar 2018.

\bibitem{Mamin1}
H.~J. Mamin, M.~Kim, M.~H. Sherwood, C.~T. Rettner, K.~Ohno, D.~D. Awschalom,
  and D.~Rugar.
\newblock {\em Science}, 339:558, 2013.

\bibitem{Kolkowitz1}
S.~Kolkowitz, Q.~P. Unterreithmeier, S.~D. Bennett, and M.~D. Lukin.
\newblock {\em Phys. Rev. Lett.}, 109:137601, 2012.

\bibitem{Zhao1}
N.~et~al. Zhao.
\newblock {\em Nat. Nanotechnol.}, 7:657 -- 662, 2012.

\bibitem{Bud1}
Arne Wickenbrock, Huijie Zheng, Lykourgos Bougas, Nathan Leefer, Samer Afach,
  Andrey Jarmola, Victor~M. Acosta, and Dmitry Budker.
\newblock {\em Applied Physics Letters}, 109:053505, 2016.

\bibitem{Bud2}
L.T. Hall, P.~Kehayias, D.A. Simpson, A.~Jarmola, A.~Stacey, D.~Budker, and
  L.C.L. Hollenberg.
\newblock {\em Nature Communications}, 7:10211, 2016.

\bibitem{Wood1}
James D.~A. Wood, David~A. Broadway, Liam~T. Hall, Alastair Stacey, David~A.
  Simpson, Jean-Philippe Tetienne, and Lloyd C.~L. Hollenberg.
\newblock {\em Physical Review B}, 94:155402, 2016.

\bibitem{Wang1}
Hai-Jing Wang, Chang~S. Shin, Scott~J. Seltzer, Claudia~E. Avalos, Alexander
  Pines, and Vikram~S. Bajaj.
\newblock {\em Nature Communications}, 2014.

\bibitem{Staudacher1}
T.~et~al. Staudacher.
\newblock {\em Science}, 339:561 -- 563, 2013.

\bibitem{Aharonovich1}
Igor Aharonovich, Andrew~D. Greentree, and Steven Prawer.
\newblock {\em Nature Photonics}, 5:397 -- 405, 2011.

\bibitem{Wyk1}
J.~H.~N. {Loubser} and J.~A. {van Wyk}.
\newblock {REVIEW: Electron spin resonance in the study of diamond}.
\newblock {\em Reports on Progress in Physics}, 41(8):1201--1248, Aug 1978.

\bibitem{Ajoy1}
M.~{Gierth}, V.~{Krespach}, A.~I. {Shames}, P.~{Raghavan}, E.~{Druga},
  N.~{Nunn}, M.~{Torelli}, R.~{Nirodi}, S.~{Le}, R.~{Zhao}, A.~{Aguilar},
  X.~{Lv}, M.~{Shen}, C.~A. {Meriles}, J.~A. {Reimer}, A.~{Zaitsev},
  A.~{Pines}, O.~{Shenderova}, and A.~{Ajoy}.
\newblock {High temperature annealing enhanced diamond 13C hyperpolarization at
  room temperature}.
\newblock {\em arXiv e-prints}, page arXiv:1911.03322, Nov 2019.

\bibitem{Shenderova1}
Laura~Dei Cas, Steven Zeldin, Nicholas Nunn, Marco Torelli, Alexander~I.
  Shames, Alexander~M. Zaitsev, and Olga Shenderova.
\newblock From fancy blue to red: Controlled production of a vibrant color
  spectrum of fluorescent diamond particles.
\newblock {\em Advanced Functional Materials}, 29(19):1808362, 2019.

\bibitem{Schwartz}
Ilai Schwartz, Jochen Scheuer, Benedikt Tratzmiller, Samuel M{\"u}ller, Qiong
  Chen, Ish Dhand, Zhen-Yu Wang, Christoph M{\"u}ller, Boris Naydenov, Fedor
  Jelezko, and Martin~B. Plenio.
\newblock Robust optical polarization of nuclear spin baths using hamiltonian
  engineering of nitrogen-vacancy center quantum dynamics.
\newblock {\em Science Advances}, 4(8):eaat8978, 2018.

\bibitem{Ajoy2}
A.~Ajoy, B.~Safvati, R.~Nazaryan, J.~T. Oon, B.~Han, P.~Raghavan, R.~Nirodi,
  A.~Aguilar, K.~Liu, X.~Cai, X.~Lv, E.~Druga, C.~Ramanathan, J.~A. Reimer,
  C.~A. Meriles, D.~Suter, and A.~Pines.
\newblock Hyperpolarized relaxometry based nuclear $t_1$ noise spectroscopy in
  diamond.
\newblock {\em Nature Communications}, 10:5160, 2019.

\bibitem{Negyedi1}
M.~Negyedi, J.~Palot{\'a}s, B.~Gy{\"u}re, S.~Dzsaber, S.~Kollarics,
  P.~Rohringer, T.~Pichler, and F.~Simon.
\newblock {\em Review of Scientific Instruments}, 88:013902, 2017.

\bibitem{Doherty1}
Marcus~W. Doherty, Neil~B. Mansonb, Paul Delaney, Fedor Jelezko, J{\"o}rg
  Wrachtrupe, and Lloyd~C.L. Hollenberg.
\newblock {\em Physics Reports}, 528:1 -- 45, 2013.

\bibitem{Hanson1}
R.~Hanson, V.~V. Dobrovitski, A.~E. Feiguin, O.~Gywat, and D.~D. Awschalom.
\newblock {\em Science}, 320:352, 2008.

\bibitem{xiao_proposal_2016}
Xing Xiao and Nan Zhao.
\newblock Proposal for observing dynamic {Jahn}-{Teller} effect by single
  solid-state defects.
\newblock {\em New Journal of Physics}, 18(10):103022, October 2016.

\bibitem{Bloembergen1}
N.~Bloembergen, S.~Shapiro, P.~S. Pershan, and J.~O. Artman.
\newblock {\em Phys. Rev.}, 114(2):445, 1959.

\bibitem{Holliday1}
Holliday K., Manson, N.~B.~Glasbeek M., and Vanoort E.
\newblock {\em J. Phys. Condes. Matter}, 1:7093 -- 7102, 1989.

\bibitem{Oort1}
van Oort~E. and Glasbeek M.
\newblock {\em Phys. Rev. B}, 40:6509 -- 6517, 1989.

\bibitem{Hiromitsu1}
Hiromitsu I., Westra J., and Glasbeek M.
\newblock {\em Phys. Rev. B}, 46:10600 -- 10612, 1992.

\bibitem{Afach1}
Samer Afach.
\newblock Results of quantum simulations of diamond to determine the
  cross-relaxation peaks of the all-optical diamond magnetometer.
\newblock {\em internal report}, 2016.

\bibitem{Auzinsh1}
M.~Auzinsh, A.~Berzins, D.~Budker, L.~Busaite, R.~Ferber, F.~Gahbauer,
  R.~Lazda, A.~Wickenbrock, and H.~Zheng.
\newblock {\em Phys. Rev. B}, 100:075204, 2019.

\bibitem{Doherty2012}
M.~W. Doherty, F.~Dolde, H.~Fedder, F.~Jelezko, J.~Wrachtrup, N.~B. Manson, and
  L.~C.~L. Hollenberg.
\newblock {\em Physical Review B}, 85(20):205203, may 2012.

\bibitem{Wrachtrup1}
P.~Neumann, R.~Kolesov, V.~Jacques, J.~Beck, J.~Tisler, A.~Batalov, L.~Rogers,
  N.~B. Manson, G.~Balasubramanian, F.~Jelezko, and J.~Wrachtrup.
\newblock Excited-state spectroscopy of single nv defects in diamond using
  optically detected magnetic resonance.
\newblock {\em New Journal of Physics}, 11:013017, 2009.

\bibitem{Bud3}
Ran Fischer, Andrey Jarmola, Pauli Kehayias, and Dmitry Budker.
\newblock {\em Physical Review B}, 87:125207, 2013.

\bibitem{hartmann_nuclear_1962}
S.~R. Hartmann and E.~L. Hahn.
\newblock Nuclear {Double} {Resonance} in the {Rotating} {Frame}.
\newblock {\em Physical Review}, 128(5):2042--2053, December 1962.
\newblock Publisher: American Physical Society.

\bibitem{rovnyak_tutorial_2008}
David Rovnyak.
\newblock Tutorial on analytic theory for cross-polarization in solid state
  {NMR}.
\newblock {\em Concepts in Magnetic Resonance Part A}, 32A(4):254--276, 2008.
\newblock \_eprint:
  https://onlinelibrary.wiley.com/doi/pdf/10.1002/cmr.a.20115.

\bibitem{Ivady}
Viktor Ivady, Krisztian Szasz, Abram~L. Falk, Paul~V. Klimov, David~J.
  Christle, Erik Janzen, Igor~A. Abrikosov, David~D. Awschalom, and Adam Gali.
\newblock Theoretical model of dynamic spin polarization of nuclei coupled to
  paramagnetic point defects in diamond and silicon carbide.
\newblock {\em Pysical Review B}, 92:115206, 2015.

\bibitem{Laraoui1}
A.~Laraoui, J.~S. Hodges, and C.~A. Meriles.
\newblock {\em Nano Lett.}, 12:3477, 2012.

\bibitem{Lange1}
G.~de~Lange, T.~van~der Sar, M.~Blok, Z.-H. Wang, V.~Dobrovitski, and
  R.~Hanson.
\newblock {\em Sci. Rep.}, 2:382, 2012.

\bibitem{Knowles1}
H.~S. Knowles, D.~M. Kara, and M.~Atatüre.
\newblock {\em Nat. Mater.}, 13:21, 2014.

\end{thebibliography}
\bibliographystyle{unsrt}

\end{document}